\documentclass[prb,preprint]{revtex4}

\usepackage{bm}
\usepackage{latexsym}
\usepackage{dcolumn}
\usepackage{amsmath,amsfonts,amssymb}
\usepackage{graphicx,epsfig}
\usepackage[caption=false]{subfig}
\usepackage{mathrsfs}
\usepackage{floatrow}

\usepackage[compact]{titlesec}
\titlespacing*{\section}
  {0pt}{2\baselineskip}{3\baselineskip}
\titlespacing*{\subsection}
  {0pt}{2\baselineskip}{3\baselineskip}

\usepackage{tikz-cd}
\usepackage{setspace}

\def\l{\left}
\def\r{\right}
\def \DM {\mathrm{d}}

\newcommand{\eq}[1]{Eq.~(\ref{#1})}
\newcommand{\delbar}{\partial\hspace*{-0.08em}\bar{}\hspace*{0.1em}}

\newcommand{\boxedeq}[2]{%
\begin{equation}
  \addtolength{\fboxsep}{5pt}
   \boxed{
   \begin{gathered}
    #1
   \end{gathered}
   }
   \label{eq:#2}
\end{equation}
}

\reversemarginpar

\begin{document}

\title{
Varying without varying: {Reparameterisations, Diffeomorphisms, General Covariance, Lie derivatives, and all that}
}

 \author{Dawood Kothawala}
 \email{dawood@iitm.ac.in}
 \affiliation{Department of Physics, Indian Institute of Technology Madras, Chennai 600 036}

\date{\today}
\begin{abstract}
\noindent
The standard way of deriving Euler-Lagrange (EL) equations given a point particle action is to vary the trajectory and set the first variation of the action to zero. However, if the action is (i) reparameterisation invariant, and (ii) generally covariant, I show that one may derive the EL equations by suitably {\it nullifying} the variation through a judicious coordinate transformation. The net result of this is that the curve remains fixed, while all other geometrical objects in the action undergo a change, given precisely by the Lie derivatives along the variation vector field. This, then, is the most direct and transparent way to elucidate the connection between general covariance, diffeomorphism invariance, and Lie derivatives, without referring to covariant derivative.
\\
\\
\noindent I highlight the geometric underpinnings and generality of above ideas by applying them to simplest of field theories, keeping the discussion at a level easily accessible to advanced undergraduates. As non-trivial applications of these ideas, I (i) derive the Geodesic Deviation Equation using first order diffeomorphisms, and (ii) demonstrate how they can highlight the connection between canonical and metric stress-energy tensors in field theories.
\end{abstract}

\maketitle
\vskip 0.5 in
\noindent
\maketitle

\begin{spacing}{0}
\tableofcontents
\end{spacing}

\section{Introduction} \label{sec:intro}
The framework of mechanics based on {\it variational principles} - a re-formulation of the mechanics of Galileo and Newton based on works of Fermat, Maupertuis, Euler, Lagrange, Hamilton, and Jacobi amongst others - is powerful not just because it reproduces, and generalises, Newton's equations of motion, but also because, regardless of what the true trajectory is, it helps us discuss the symmetries and invariances of the physical system under consideration. ``{\it Nature is thrifty in all its actions}", wrote Maupertuis, thereby capturing the essence of the so called action principle \cite{goldstein-book, sudarshan-mukunda-book, ajp-taylor}, which essentially states that the trajectory which solves Newtons equations of motion is the one for which the value of a certain quantity - the {\it action} -  is stationary; this quantity is often taken to be of the form
$
\mathcal A\l[x\r] = \int \limits_{\lambda_{\rm I}}^{\lambda_{\rm F}} L(x^i, \dot x^i, \lambda) \, \DM \lambda
$
, where $\lambda$ is some parameter along the path, $\dot{x}^i = \DM x^i/\DM \lambda$, and $L$ is the Lagrangian function. The object $\mathcal A\l[x\r]$ depends on the entire function (curve) $x^i(\lambda)$, and is called a {\it functional}. In classical mechanics, one calls $\mathcal A\l[x\r]$ an action, and often chooses the parameter $\lambda$ as time $t$, but of course, one does not have to. The choice of parameter - or rather, the {\it irrelevance} of this choice - will be one of our main points of focus in the next section. We will then take this as a motivating example to explain more advanced notions of diffeomorphism invariance and Lie derivatives, by connecting them to general covariance, which is essentially the {\it irrelevance} of choice of coordinates, no different from the above mentioned {\it reparameterization invariance} of the point particle action.

The main motivation behind this paper is to provide a robust and transparent analysis of what one {\it really means} by diffeomorphism invariance, how it is related to general covariance, and what do Lie derivatives have to do with any of this. Such ideas are usually either discussed in terms of purely mathematical definitions in differential geometry, or presented using words such as ``\ldots {\it dragging a tensor along a vector field} \ldots" in books on mathematical physics or general relativity. More often than not, this leaves one with a discomforting feeling of having not really understood the key insight that diffeomorphism invariance is supposed to convey. Perhaps someone encountering these concepts for the very first time, and struggling to really understand their significance, would feel a bit more comforted when (s)he realises that Einstein himself struggled with the issue of general covariance, which he articulated in terms of his famous {\it Hole argument} \cite{hole-argument}. Decades of discussions on this argument has led to several insights, but the most powerful one is that physical {\it events} (such as a intersection of trajectories of two observers) are more fundamental than spacetime {\it points}. Indeed, the careful reader will find the theme ``coordinates are mere labels" recurring throughout this paper, and I would encourage him/her to connect this with the essence of the hole argument.

It is my hope that describing clearly notions such as reparameterization invariance, general covariance, diffeomorphism invariance and Lie derivatives, and their inter-connections, will help appreciate the finer points and subtleties associated with these ideas. Moreover, it becomes clear that much of this discussion can be done before one learns about {\it covariant derivatives}, thereby making it clear that, unlike covariant derivatives, Lie derivatives do not require any additional structure on spacetime.

Here is a brief summary of the paper: In {\bf Sec II}, I discuss the familiar case of point particle action with a special focus on its properties under change of parameterization of the curves, and introduce the concept of variation without variation. {\bf Sec III} then moves on from point-particles to fields, and the main focus is to highlight the interconnection between diffeomorphims and choice of coordinates by borrowing the key ideas from {\bf Sec II}. Finally, {\bf Sec IV} discusses two examples which are intended to serve as non-trivial applications of the idea of varying without varying.

{\it Notations}: I will use mostly use notations and conventions of \cite{mtw-book}, and would also refer the reader to the excellent primer on General Relativity by Price \cite{ajp-price-gr-primer}. To denote fields such as $\phi(t,x,y,z)$, I will often use $\phi(x^i)$ or $\phi(x)$, depending on notational convenience. Also, $\partial_k \phi \equiv \partial \phi/\partial x^k$, and I use the notation $t'^a(x')$ to denote the components of a vector(-field) $t^a(x)$ in coordinates $x'$. Standard summation convention has been employed, in which repeated indices are assumed to be summed over all possible values.

\section{Point particle action - {\it Redux}}
\begin{figure}[!htb]
\begin{center}
\scalebox{0.3}{\includegraphics{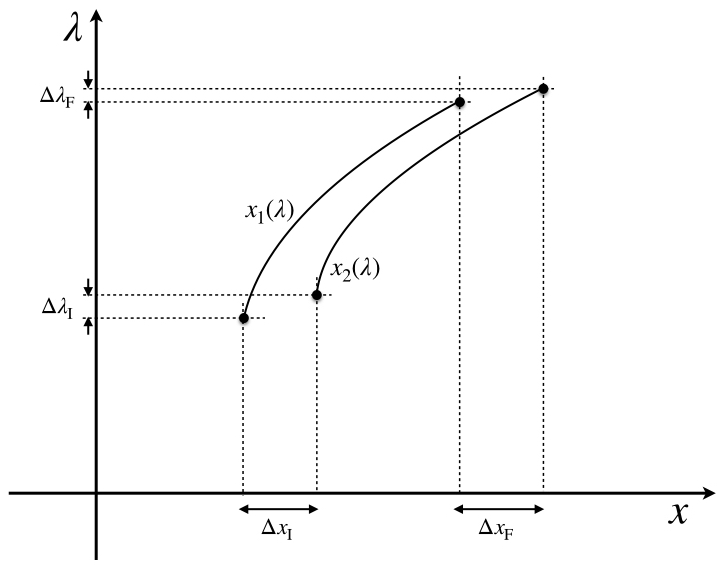}} 
\caption{General variation of a curve (depicted here in a one dimensional space). The difference in the corresponding actions depends on the boundary variations highighted in the figure (see text).}
\label{fig:action-full-variation}
\end{center}
\end{figure}

Given an action, the basic question of interest is the following: given two curves $x_1(\lambda)$ and $x_2(\lambda)$, what is the first order difference in their actions? This can be easily derived along the lines done in standard courses, except that we will here keep the boundary points also different (see \textbf{Fig.\ref{fig:action-full-variation}}).

\begin{eqnarray}
\delta \mathcal A 
&=& \int \limits_{\lambda_{\rm I}}^{\lambda_{\rm F}} \mathscr{E}_i \, \delta x^i \, \DM \lambda + \Biggl[ p_i \Delta x^i - E \Delta \lambda \Biggl]_{\rm I}^{\rm F}
\label{eq:action-total-variation}
\end{eqnarray}
where $\delta x^i = x^i_2(\lambda) - x^i_1(\lambda)$, and we have used the usual definitions
	\begin{eqnarray}
	p_i &=& \partial L/\partial \dot x^i \hspace{1.95cm} \mathrm{(Conjugate \; momentum)}
	\nonumber \\
	E &=& \dot x^i \partial L/\partial \dot x^i - L \hspace{0.85cm} \mathrm{(Energy \; function)}
	\nonumber \\
	\mathscr{E}_i &=& \partial L/\partial x^i - \dot p_i  \hspace{1.2cm} \mathrm{(Euler-Lagrange \; function(al))}
	\nonumber
	\end{eqnarray}

The above result is readily derived as follows (see also \cite{mtw-action}): 
	\begin{eqnarray}
	\delta \mathcal A &=& {\mathcal A\l[x^i_2\r]} - {\mathcal A\l[x^i_1\r]}
	\nonumber \\
	&=& 
	\int \limits_{\lambda_{\rm I} + \Delta \lambda_{\rm I}}^{\lambda_{\rm F} + \Delta \lambda_{\rm F}} L(x^i_2, \dot x^i_2, \lambda) \, \DM \lambda - \int 	
	\limits_{\lambda_{\rm I}}^{\lambda_{\rm F}} L(x^i_1, \dot x^i_1, \lambda) \, \DM \lambda
	\nonumber \\
	&=& 
	\int \limits_{\lambda_{\rm I}}^{\lambda_{\rm F}} \Biggl( L(x^i_2, \dot x^i_2, \lambda) - L(x^i_1, \dot x^i_1, \lambda) \Biggl) \DM \lambda + \left[ L \Delta \lambda 
	\right]_{\rm I}^{\rm F}
	\nonumber \\
	&=& \int \limits_{\lambda_{\rm I}}^{\lambda_{\rm F}} \mathscr{E}_i \, \delta x^i \, \DM \lambda + \left[ p_i \delta x^i + L \Delta \lambda \right]_{\rm I}^{\rm F}
	\label{eq:action-total-variation-steps}
	\end{eqnarray}
	where we have used Lebniz's Integral Rule (differentiation under the integral sign) in the third step.
This immediately yields Eq.~(\ref{eq:action-total-variation}) once we replace the {\it local} variations $\delta x^i$ in terms of the so called {\it total} variations 
$\Delta x^i(\lambda) := x^i_2(\lambda+\Delta \lambda) - x^i_1(\lambda)$, as follows
	\begin{eqnarray} 
	\Delta x^i({\lambda}) 
	&=& 
	\underbrace{\l\{ x^i_2(\lambda+\Delta \lambda)  - x^i_2(\lambda) \r\} }_{\dot x^i \Delta \lambda}
	+ 
	\underbrace{ \l\{ x^i_2(\lambda)  - x^i_1(\lambda) \r\} }_{\delta x^i(\lambda)}
	\nonumber
	\end{eqnarray} 
Hence, $\delta x^i = \Delta x^i - \dot x^i \Delta \lambda$. The result in Eq.~(\ref{eq:action-total-variation}) is very important, since we will see that a very similar structure emerges also in the case of more complicated field theory actions, going well beyond the point particle case.

\subsection{Reparameterization invariant point particle action} \label{sec:reparam}
What does it mean for this action to be reparameterization invariant? It means the following: 

{\it Reparameterization invariance}: If one chooses some parameter other than $\lambda$, then, for the new trajectory obtained by {\it assuming that $x^i(\lambda)$ transforms as a scalar under the reparameterization}, the action does not change. 

The italicised part above is particularly important when one relates this invariance with the vanishing of the Hamiltonian for such systems, as we shall do below \cite{henneaux-teitelboim}. To clearly see what it means, consider the infinitesimal reparameterizations: $\lambda \rightarrow \lambda' = \lambda + \epsilon f(\lambda)$. What we mean by the scalar transformation property of trajectories is that the trajectory $x^i_1(\lambda)$ transforms to the new trajectory $x^i_2(\lambda)$ given by 
\begin{eqnarray}
x^i_2(\lambda+\epsilon f(\lambda)) &=& x^i_1(\lambda)
\\ 
x^i_2(\lambda) &=& x^i_1(\lambda) - \epsilon f(\lambda) \dot x^i(\lambda)
\end{eqnarray}
keeping everything to first order. Note that the scalar transformation property of $x^i(\lambda)$ (the first equality above) immediately implies $\Delta x^i=0$. What it instructs is the following: take a point $P$ on the original trajectory, and drag it to the parameter value $\lambda+\epsilon f(\lambda)$ keeping the value of $x^i(\lambda)$ the same. What we have essentially done is capture the effects of change of parameter in the description of a path as a specific variation of that path.

%
This requirement corresponds to $\Delta \lambda=\epsilon f(\lambda)$, $\delta x^i = - \epsilon f \dot x^i$ and, of course, $\Delta x^i=0$. Plugging this in Eq.~(\ref{eq:action-total-variation})
\begin{eqnarray}
\delta \mathcal A &=& -\epsilon \int \limits_{\lambda_{\rm I}}^{\lambda_{\rm F}} \mathscr{E}_i \, f \dot x^i \, \DM \lambda - \epsilon \l[ E f \r]_{\rm I}^{\rm F}
\nonumber \\
&=& -\epsilon \int \Biggl[ f \l( \dot x^i \frac{\partial L}{\partial x^i} - \dot x^i \frac{\DM p_i}{\DM \lambda} \r) + \frac{\DM (E f)}{\DM \lambda} \Biggl] \DM \lambda
\nonumber \\
&=& -\epsilon \int \Biggl[ f \l( \dot x^i \frac{\partial L}{\partial x^i} + \ddot x^i \frac{\partial L}{\partial \dot x^i} - \frac{\DM (p_i \dot x^i)}{\DM \lambda} \r) + \frac{\DM (E f)}{\DM \lambda} \Biggl] \DM \lambda
\nonumber \\
&=& -\epsilon \int \Biggl[ f \l( \frac{\DM L}{\DM \lambda} - \frac{\partial L}{\partial \lambda} - \frac{\DM (p_i \dot x^i)}{\DM \lambda} \r) + \frac{\DM (E f)}{\DM \lambda} \Biggl] \DM \lambda
\nonumber \\
&=& -\epsilon \int \Biggl[ f \l( - \frac{\DM E}{\DM \lambda} - \frac{\partial L}{\partial \lambda} \r) +  \frac{\DM (E f)}{\DM \lambda}  \Biggl] \DM \lambda
\nonumber \\
&=& +\epsilon \int \left( \frac{\partial L}{\partial \lambda} f - E \dot f \right) \DM \lambda
\label{Eeq0-deriv}
\end{eqnarray}
where in the 4th equality we have used the total derivative of $L(x^i, \dot x^i, \lambda)$:
$$
\frac{\DM L}{\DM \lambda} = \frac{\partial L}{\partial \lambda} + \dot x^i \frac{\partial L}{\partial x^i} + \ddot x^i \frac{\partial L}{\partial \dot x^i}
$$ 

The above expression for $\delta \mathcal A$ can now be used to deduce the consequences of reparameterization invariance of the action. {\it If} an action is invariant under arbitrary reparameterizations (that is, for arbitrary $f(\lambda)$ and hence $\dot f(\lambda)$), then $\delta \mathcal A=0$ and we must have
\boxedeq{
E=0 \hspace{.5cm} {\rm and}  \hspace{.5cm}  \frac{\partial L}{\partial \lambda} =0
}{Eeq0}
(The reader should be able to convince himself/herself of this by a familiar argument: Choose a reparameterization $f=$ constant to deduce ${\partial L}/{\partial \lambda} =0$, and plug this back in for an arbitrary reparametrization with $\dot f \neq 0$ to deduce $E=0$.)
If one can invert $\dot x^i$ in terms of $p_i$, then the first condition above is the well known condition of vanishing of the Hamiltonian for a reparameterization invariant action. Using the definition of $E$, the above condition implies that, for a reparameterization invariant lagrangian, we must have 
\begin{eqnarray}
\dot x^i \frac{\partial L}{\partial \dot x^i} = L
\end{eqnarray}
If one insists the lagrangian $L$ to be quadratic in velocities, it is easy to see that the only solution of the above equation is {\it square root} lagrangian (familiar from Special and General Relativity)
$$
L = - \sqrt{- g_{ab} t^a t^b} \;\;\; ; \;\;\; t^a = \frac{\DM x^a(\lambda)}{\DM \lambda}
$$
where $g_{ab}(x)$ is, for now, as arbitrary matrix whose components may depend on coordinates $x^i$ (as we will see shortly, $g_{ab}(x)$ is the metric tensor). We will discuss this example in more detail in the next subsection, \ref{subsec:var-without-var}. 

The second condition, ${\partial L}/{\partial \lambda}=0$, is intuitively obvious: it simply says that for an action to be invariant under reparameterizations, the Lagrangian should not have any explicit dependence on the parameter. 

We may note that we have assumed nothing whatsoever about the trajectory being on-shell, that is, $x^i(\lambda)$ need not satisfy the equations of motion. Each of the above steps for point particle action and its variation will have a counterpart in more complicated actions one encounters in field theories, with very similar structures as above emerging while discussing {\it diffeomorphism invariance} in such theories. Indeed, it is one of the aims of this paper to highlight the common geometrical aspects associated with invariances of an action under {\it reparameterizations} of the variable(s) over which one integrates a Lagrangian to obtain the action.

\subsection{``Variation without variation"} \label{subsec:var-without-var}
Having discussed consequences of reparameterization invariance, we now examine the consequence of the fact that our actions are coordinate invariant, and use this general covariance to employ the following {\it trick} \cite{ajp-diff-g}:

\textit{In the variation $\delta \mathcal A = {\mathcal A\l[x^i_2\r]} - {\mathcal A\l[x^i_1\r]}$, evaluate ${\mathcal A\l[x^i_2\r]}$ in a different coordinate system, one in which 
$x^i_2$ ``looks the same" as $x^i_1$.}

\begin{figure*}
\floatbox[{\capbeside\thisfloatsetup{capbesideposition={right,top},capbesidewidth=4cm}}]{figure}[\FBwidth]
{\caption{This figure demonstrates the difference between standard method of variation of a point-particle action (top) and the version in which one effectively does not vary the path at all, but instead performs a suitable coordinate transformation on one of the paths, here ${\mathscr C}_2$, so that the paths ``look the same", albeit in different coordinate systems. This "variation without variation" naturally introduces Lie derivatives when computing the variation of the actions. See text for details.
}\label{fig:active-passive-point-particle}}
{\includegraphics[width=0.7\textwidth]{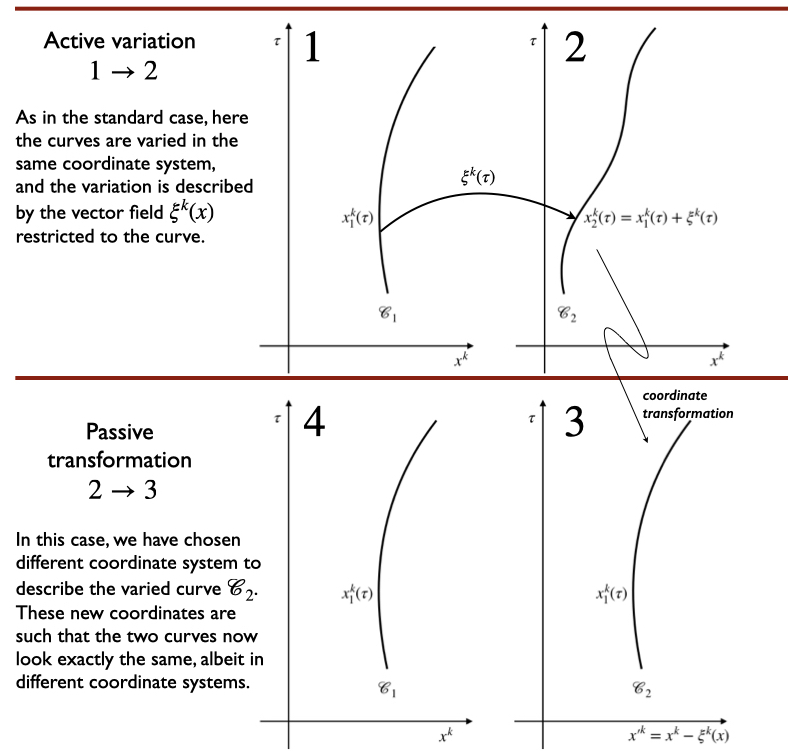}}
\end{figure*}

The procedure is trivial, and is illustrated clearly in \textbf{Fig.\ref{fig:active-passive-point-particle}}. Because of covariance, we should get the same result as before, although now the curves are not being varied at all! We are thus ``varying without varying", to use a Wheeler{\it esque} phrase. Of course, something must change, since we 
know $\delta \mathcal A$ is non-zero in general. To see what this is, we need to be more specific about the form of the action. We will assume that the action can be written in the following form:
\begin{eqnarray}
\mathcal A\l[x\r] &=& \int \limits_{\lambda_{\rm I}}^{\lambda_{\rm F}} L(x^i, \dot x^i, \lambda) \, \DM \lambda
\equiv \int \limits_{\lambda_{\rm I}}^{\lambda_{\rm F}} L( g_{ab}(x^i), t^i, \lambda) \, \DM \lambda
\end{eqnarray}
where $t^i=\dot x^i$ is the tangent vector to the curve. That is, we assume that there exists an object $g_{ab}(x^i)$ - technically a rank two tensor - such that all the explicit coordinate dependence in the action can be absorbed in $g_{ab}(x^i)$. Given a suitable (tensor) transformation property for $g_{ab}(x^i)$, one can then construct lagrangian which is a scalar. Had we allowed an explicit dependence on coordinates, $L$ would in general not have the same form when transformed from one coordinate system to another. The general condition for any such lagrangian to be form invariant is that it depends on coordinates only through geometric quantities which transform as tensors under change of coordinate system. Although plausible, one can give a mathematically rigorous proof of this fact \cite{iyer-wald}. We will demonstrate this explicitly in the context of a scalar field theory in {\bf Sec IV (B)}.
%

The transformation properties of $g_{ab}$ and $t^i$ will now ensure that $L$ is a scalar. At this point, we urge the reader to have a look at Appendix \ref{app:Lie-derivs} on Lie derivatives to see their definition; as we will see below, they appear naturally in the derivation variation of the action with the curve fixed.

We now come to the slick, yet perhaps the trickiest, part of the computation. We start at the 3rd step of Eq.~(\ref{eq:action-total-variation-steps}), in our new notation stated above:
\begin{eqnarray}
\delta \mathcal A &=&  
\underbrace{\int \limits_{\lambda_{\rm I}}^{\lambda_{\rm F}} L(g_{ab}(x_2^i), t_2^i, \lambda) \DM \lambda}_{\mathrm{change~coordinates:~} x'^k = x^k - \epsilon \xi^k(x)} 
\;\;\;\;\;\; - \;\; 
\underbrace{\int \limits_{\lambda_{\rm I}}^{\lambda_{\rm F}} L(g_{ab}(x_1^i), t_1^i, \lambda) \DM \lambda}_{\mathrm{keep~coordinates~unchanged}}   \;\; + \;\;  \left[ L \Delta \lambda \right]_{\rm I}^{\rm F}
\nonumber \\
&=&  
\int \limits_{\lambda_{\rm I}}^{\lambda_{\rm F}} L(g'_{ab}({x'}_2^i), {t'}_2^i, \lambda) \DM \lambda 
\;\;\; - \;\;  \int \limits_{\lambda_{\rm I}}^{\lambda_{\rm F}} L(g_{ab}(x_1^i), t_1^i, \lambda) \DM \lambda   \;\; + \;\;  \left[ L \Delta \lambda \right]_{\rm I}^{\rm F}
\nonumber \\
&=&  
\int \limits_{\lambda_{\rm I}}^{\lambda_{\rm F}} L(g'_{ab}({x}_1^i), {t}_1^i, \lambda) \DM \lambda 
\;\;\;\;\; - \;\;  \int \limits_{\lambda_{\rm I}}^{\lambda_{\rm F}} L(g_{ab}(x_1^i), t_1^i, \lambda) \DM \lambda   \;\; + \;\;  \left[ L \Delta \lambda \right]_{\rm I}^{\rm F}
\nonumber \\
&=&  
\int \limits_{\lambda_{\rm I}}^{\lambda_{\rm F}} \Biggl[ L({g}_{ab} + \mathscr{L}_{\bm \xi} g_{ab}, {t}_1^i, \lambda) \;-\; L(g_{ab}, t_1^i, \lambda) \Biggl] \DM \lambda   \;\; + \;\;  \left[ L \Delta \lambda \right]_{\rm I}^{\rm F}
\nonumber \\
&=&  
\int \limits_{\lambda_{\rm I}}^{\lambda_{\rm F}} \l[ \frac{\partial L}{\partial g_{ab}} \mathscr{L}_{\bm \xi} g_{ab} \r] \DM \lambda   \;\; + \;\; \left[ L \Delta \lambda \right]_{\rm I}^{\rm F}
\label{eq:action-total-variation-active}
\end{eqnarray}
Let me now explain the crucial steps above: 

{\bf 2nd equality}: As suggested in \textbf{Fig.\ref{fig:active-passive-point-particle}}, we implement the coordinate transformation that effectively maps the shifted curve to the same curve.

{\bf 3rd equality}: {\it The most important step!} Since our coordinate transformation was tailor-made to make the shifted curve $\mathscr{C}_2$ look exactly like $\mathscr{C}_1$ in the new coordinates, we have 
$$g'_{ab}({x'}_2^i) \equiv g'_{ab}({x}_1^i) \;\;\;\; ; \;\;\;\; {t'}_2^i \equiv {t}_1^i$$
Note that the net effect of all this is simply to replace 
$$g_{ab}(\cdot) \to g'_{ab}(\cdot) $$
in the action for the second curve, while keeping the curve itself fixed.

{\bf 4th equality}: Observe that now, the difference between the metrics appearing in the two terms in the 3rd equality is nothing but a Lie derivative (see Appendix \ref{app:Lie-derivs}). Notice that the Lie derivative now appears naturally as a consequence of diffeomorphism annulled via a coordinate transformation. Another consequence of this same fact is that, with tangent vectors matched, we have
$${t'}_2^i(x_1) - {t}_1^i(x_1) =0 
$$ 
(note that everything is now evaluated on the curve $\mathscr{C}_1$ given by $x^i_1(\lambda)$). The above condition can be given an elegant geometric intepretation in terms of Lie derivative, as follows: One may characterise the variation of a curve with tangent $\bm t(\lambda)$ by defining a one-parameter family of curves with tangents $\bm T(\lambda; \alpha)$, $\alpha \in \mathbb{R}$, with $\bm T(\lambda; 0)=\bm t(\lambda)$ being the base (original) curve. Similarly, $\bm T(\lambda; \alpha')=\bm t'(\lambda)$ for the varied curve. When one varies a curve along a vector field $\xi^i(x)$ (as above), $\alpha$ can be chosen as the parameter along $\xi^i(x)$. Then, the difference 
${t'}_2^i(x) - {t}_1^i(x)$ is easily seen to be (apart from a minus sign) the same as the Lie derivative $\mathcal{L}_{\bm \xi} \bm T(\lambda; \alpha)|_{\alpha=0}$ (see Eq.~(\ref{eq:lie-deriv-vec-field}) of Appendix \ref{app:Lie-derivs}) . Therefore, the condition above is equivalent to $\mathcal{L}_{\bm \xi} \bm T(\lambda; \alpha)|_{\alpha=0}=0$. To avoid notational clutter, we will simply write $\mathcal{L}_{\bm \xi} \bm T(\lambda; \alpha)|_{\alpha=0} \overset{\rm def}{:=} \mathcal{L}_{\bm \xi} \bm t$, which is what we will mean by the symbol $\mathcal{L}_{\bm \xi} \bm t$ (it should not be misinterpreted as applicable to a single tangent vector, but rather to tangent vectors associated with the one-parameter class of curves being varied). The vanishing of Lie derivative of one vector field with respect to another has a very elegant geometrical interpretation in terms of closure of an infinitesimal rectangle formed from $\bm \xi$ and $\bm t$; I do not discuss this aspect here, one place worth checking would be \cite{mtw-book}. The astute reader might be wondering: What is the analog of this condition in the standard variational principle for curves one encounters in introductory discussions? Well, the answer is hidden rather subtly in the assumption one often makes in the standard variation principle: $\delta\left({dx^i}/{d\lambda}\right) = {d}(\delta x^i)/{d\lambda}$. This is not unrelated to the comment just made concerning the geometrical meaning of the vanishing of Lie derivative of two vector fields. I leave it as an exercise to the reader to connect these two using elementary calculus. 

\textbf{\textit{Example}:} To connect the above form with Eq.~(\ref{eq:action-total-variation}), we need more specific information about the Lagrangian\footnote{We may in fact proceed a bit keeping things more general, but that to me does not seem to be of much pedagogical value.}. Let us consider, then, the Lagrangian
$$
L = - \sqrt{- g_{ab} t^a t^b} \;\;\; ; \;\;\; t^a = \frac{\DM x^a(\lambda)}{\DM \lambda}
$$
familiar from Special and General Relativity. This is essentially the Lagrangian of a relativistic point particle (with mass $m$ set to unity). We have
$$
 \frac{\partial L}{\partial g_{ab}} = - \frac{1}{2L} t^a t^b \;\;\; ; \;\;\; p_a = \frac{\partial L}{\partial t^a} = - L^{-1} g_{ab} t^b \;\;\; ; \;\;\; E = 0
$$
A few steps of algebra then give
\begin{widetext}
\begin{eqnarray}
\frac{\partial L}{\partial g_{ab}} \mathscr{L}_{\bm \xi} g_{ab} &=& \frac{1}{L} g_{ij} \xi^j \Biggl[ 
\underbrace{ t^a \partial_a t^i + 
\overbrace{ \frac{1}{2} g^{im}\l( - \partial_m g_{bc} + \partial_b g_{cm} + \partial_c g_{mb} \r) }^{ := \; \Gamma^i_{\phantom{i}bc} \; \equiv \; {\rm Christoffel \; symbols}} 
t^b t^c}_{= \; a^i \; \equiv \; {\rm covariant \; acceleration}} 
\Biggl] - \frac{1}{L} t^a \partial_a \l( g_{ij} t^i \xi^j \r)
\nonumber \\
&=& - g_{ab} \xi^a a^b + \frac{\DM}{\DM \lambda} \l( g_{ab} t^a \xi^b \r)
\label{eq:geodesic-eq-active}
\end{eqnarray}
\end{widetext}
where in the last step we have put $L=-1$, which is the standard textbook choice of parametrization in relativity. We have also introduced the acceleration vector $a^i$ 
associated with the curve
\begin{eqnarray}
a^i = t^a \partial_a t^i + \Gamma^i_{\phantom{i}bc} \, t^b t^c = \frac{\DM t^i}{\DM \lambda} + \Gamma^i_{\phantom{i}bc} \, t^b t^c
\label{eq:geodesic-eq}
\end{eqnarray}
It is now a trivial and quick exercise to show that Eq.~(\ref{eq:geodesic-eq-active}), when plugged into Eq.~(\ref{eq:action-total-variation-active}), reproduces Eq.~(\ref{eq:action-total-variation}). 

\textit{Geodesics}: It follows from above that, the Euler-Lagrange equations for a free particle with lagragian $L = - \sqrt{- g_{ab} t^a t^b}$ corresponds to $a^i = 0$. Curves which satisfy this equation are known as {\it geodesic} curves. Quite generically, geodesic curves are curves of extremal length between two points, assuming that a unique such curve exists. This follows from the fact that the quantity $\int L \DM \lambda = - \int \sqrt{- g_{ab} \DM x^a \DM x^b}$ is nothing but (minus) the (squared) length between the two fixed points in spacetime. Of course, the same discussion works in ordinary space as well, with $L = \sqrt{+ g_{ab} t^a t^b}$.

\textit{Killing vectors}: It is worthwhile to take a brief digression here and use the above example, along with the expressions derived in this section, to discuss conservation laws associated with special class of vector fields $\xi^i(x^k)$ that capture some symmetries of the system. This is easily done, starting with Eq.~(\ref{eq:action-total-variation-active}) with $\Delta \lambda=0$. Suppose the variation vector fields $\xi^i(x^k)$ are such that $\delta \mathcal A=0$, this would immediately imply $({\partial L}/{\partial g_{ab}}) \mathscr{L}_{\bm \xi} g_{ab}=0$ and hence $\mathscr{L}_{\bm \xi} g_{ab}=0$. From Eq.~(\ref{eq:geodesic-eq-active}), this implies $\frac{\DM}{\DM \lambda} \l( g_{ab} t^a \xi^b \r)= g_{ab} \xi^a a^b$. When equations of motion are satisfied (that is, when $a^i=0$), this statement implies that the quantity $g_{ab} t^a \xi^b$ will be a constant. Given a metric, one can look at the condition $\mathscr{L}_{\bm \xi} g_{ab}=0$ as providing differential equations to determine the vector fields $\xi^i(x^k)$. Such vector fields are known as Killing vectors, and they are immensely helpful in solving equations of motion since they provide useful first integrals of these equations. If there are additional fields (other than $g_{ab}$) in the lagrangian, the invariance of the action will require vanishing of Lie derivatives of these additional fields along $\xi^i(x^k)$ as well.  

\textit{Geometric interpretation of the Lie derivative}: Following carefully through the steps above reveal the geometric significance of the Lie derivative, the definition for which is given in Appendix \ref{app:Lie-derivs}. At this point, it might help the reader to have a look at the last paragraph in that appendix. As stated there there, Lie derivatives provide a notion of derivative in which the change in an arbitrary tensor field, symbolically denoted $\mathbb T$, as one goes from a point $p$ to $q$ along a given vector field $\xi^k$ is evaluated as follows: (i) make a coordinate transformation such that \textit{numerical values of coordinates} of $q$ in the new coordinate system are same as those of $p$ in the old coordinate system, (ii) set up a new tensor field $\mathbb T_{\rm new}(q)$ at $q$ such that it's {\it components} in the new coordinate system have same \textit{numerical values} as the corresponding components of $\mathbb T$ had {\it at $p$}, and finally (iii) transform $\mathbb T_{\rm new}(q)$ back to the original coordinate system and take it's difference with $\mathbb T(q)$. This difference is the Lie derivative. One may have to read these steps a couple of times for them to sink in, and to verify that these reproduce the expression for Lie derivative quoted in Appendix \ref{app:Lie-derivs}.

Our discussion in this section fills an important gap that exists between the above, admittedly obscure, definition of Lie derivative and the real geometric reason as to why it is physically relevant, along with answering the question: why/how is a Lie derivative connected with general covariance and diffeomorphism invariance. The rest of the paper will illuminate on this connection further.

\section{Diffeomorphism Invariance and General covariance}

\begin{figure*}
    {{\includegraphics[width=0.975\textwidth]{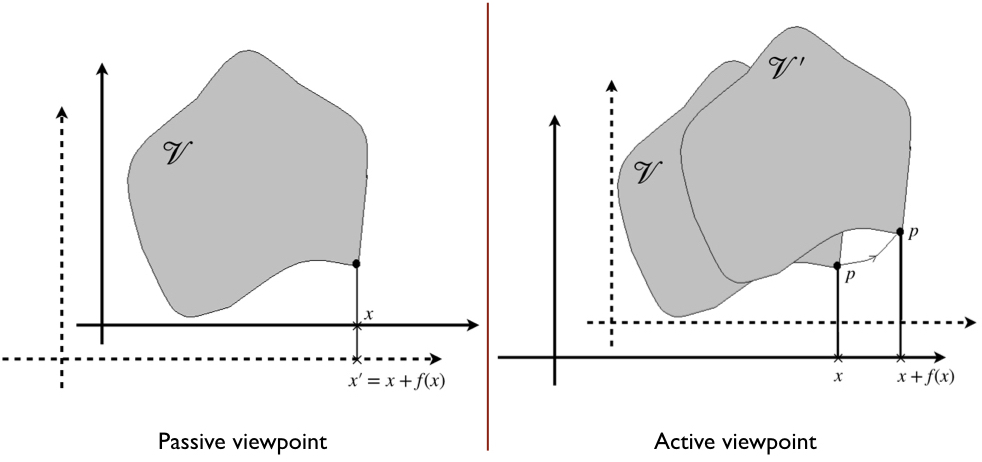} }}%
    \caption{
    {\it Left}: The {passive} viewpoint, in which the same region $\mathcal{V}$ of spacetime is described in two different coordinate systems, $x'^k$ (dashed 
    axes) and $x^k$.
    {\it Right}: The {active} viewpoint, in which an infinitesimal diffeomorphism moves the point along the flow of some vector field $\xi^k(x)$, thereby mapping $\mathcal{V} 
    \to \mathcal{V'}$. One then judiciously uses a different coordinate system $x'^k$ (dashed axes) to describe $\mathcal{V'}$, by demanding that the new coordinates of 
    the shifted point are numerically equal to those of the unshifted point in original coordinates $x^k$.
    }%
    \label{fig:set1}%
\end{figure*}

One can often analyse the symmetries of a system by looking at the invariance of the action under particular transformations - this is essentially the content of Noether's theorems. However, much of the applications of Noether's theorem discussed in elementary courses often focus on deriving identities/conservation laws that hold {\it when the equations of motion are satisfed} \cite{noether-thm}. To do this, we simply consider Eq.~(\ref{eq:action-total-variation}) for paths for which equations of motion hold, that is, $\mathscr{E}_i=0$, and hence:
$$
\underbrace{\delta \mathcal A = 0}_{\rm invariance} = \Biggl[ p_i \Delta x^i - E \Delta \lambda \Biggl]_{\rm I}^{\rm F}
$$
from which one may immediately derive conservation laws under specific transformations $\Delta x^i, \Delta \lambda$. (Notice that $\delta \mathcal A = 0$ above due to some symmetry, and this has nothing to do with the action principle, in which the action is varied with fixed end-points.)

However, what is not often emphasised, at least at an undergraduate level, is the fact that we can also get knowledge of certain identities/conservation laws satisfied by the dynamical variables in the action {\it which hold regardless of whether these dynamical variables satisfy the equations of motion or not}.  

This latter can be done simply by formulating the action in a particular manner, and we have already demonstrated this in the case of point particle action in Section \ref{sec:reparam}. The reader will recall that nowhere in that discussion did we use the equations of motion. I will now take up the issue of what is the analog of such results in field theories, focussing on two questions:
\begin{enumerate}
\item What is the analog of identities such as $\partial L/\partial \lambda$ and $E=0$ in case of field theories? 
\item What is the connection between General Covariance (invariance under coordinate transformations), Active diffeomorphims (mappings of points/events in the same space(time)), and Lie derivatives.
\end{enumerate}

By General Covariance, one means formulating the laws of physics (action, or equations of motion) in a manner such that they have the same form in any coordinate system. This, of course, is a matter of choice. One may formulate, for example, the first of Maxwell's equations either as $\bm \nabla \bm \cdot \bm E = \rho/\epsilon_0$, or write it as $\partial_x E^x + \partial_y E^y + \partial_z E^z = \rho/\epsilon_0$ in Cartesian coordinate system. The former (vector) version looks the same in any other coordinate system, while the latter one will look different when written, for example, in spherical polar coordinates (the equation does {\it not} become $\partial_r E^r + \partial_\theta E^\theta + \partial_\phi E^\phi = \rho/\epsilon_0$). Needless to say, we prefer the form $\bm \nabla \bm \cdot \bm E = \rho/\epsilon_0$ since it is generally covariant, and looks the same in any coordinate system. The above discussion is applicable to any set of laws describing any of the known physical interactions. Why, then, do we hear about general covariance primarily in the context of gravity? The importance of general covariance in Einstein's formulation of general relativity arises because of Einstein's assertion that all the effects of gravity on a physical system can be taken into account by writing the equations of motion of that system in flat spacetime in a generally covariant manner, and then simply demanding these equations remain true in presence of gravity. The gravitational effects are taken into account through the metric tensor $g_{ab}$. In the above example of Maxwell's equations, this would amount to describing $\bm E$ and $\bm B$ in terms of a four dimensional tensorial object, and elevate the three dimensional $\bm \nabla$ and the dot products to their suitable four dimensional counterparts using the metric $g_{ab}$. The principle of general covariance then tells you that the equations you obtain by doing so describe the electromagnetic field in an arbitrary gravitational background! And it is this statement that carries the physical content of the principle of general covariance. The above example should also make it clear that the principle of general covariance, although simple, is very restrictive as it says nothing about all possible curvature terms that might appear in the equations, since these terms will be zero in flat spacetime. Let us now come to diffeomorphism invariance. Active diffeomorphism is a map from a manifold to itself, such that the map and its inverse are differentiable. Such a map can be thought of as moving around points in the same manifold. A priori, this is a very different action that changing coordinate charts on the same manifold, and hence unrelated to general covariance. However, one can show that a theory formulated in a generally covariant manner will also be invariant under diffeomorphims, provided all the geometric quantities appearing in the action are acted upon by these diffeomorphisms. Highlighting this connection is one of the aims of this paper. For a very illuminating discussion of the differences between general covariance (sometimes also referred to as passive diffeomorphims) and active diffeomorphisms, we refer the reader to the book by Rovelli \cite{rovelli}. Another useful reference that addresses the connection between general covariance and diffeomorphism invariance is the article by Norton \cite{norton}.

We will use the simplest of field theories - the scalar field theory - to make our point, and leave as a guided homework generalisations to more complicated field theories.

What we have in mind is a scalar field theory described by an action
\begin{eqnarray}
\mathcal A\l[\phi\r] &=& \int \limits_{\{\mathcal V, D\}} \mathcal{L}(\phi, \partial_k \phi, g_{ab}, x^i) \, \sqrt{-g} \,  \DM^4 x
\nonumber \\
&=& \int \limits_{\{\mathcal V, D\}} L(\phi, \partial_k \phi, g_{ab}, x^i) \, \DM^4 x
\end{eqnarray}
where $\mathcal{L}$ is the Lagrangian density describing the scalar field $\phi(x)$, and we have defined $L =  \mathcal{L} \sqrt{-g}$ for future convenience. The cognoscenti will recognise $\sqrt{-g} \DM^4 x$ as the coordinate invariant volume element.

Let us know try to understand the connection between General covariance and Diffeomorphism invariance, {\it a la} reparameterization invariance which we discussed for point particle action.

\subsection{The Passive version: General covariance}

Start with the admittedly trivial fact that a generally covariant action has same values in different coordinate systems, something that we are used to from basic courses. One can solve a problem in polar or cartesian coordinates, depending on convenience, and physical results do not depend on such choice of coordinates. 

Consider, then, a particular physical region of spacetime, call it $\mathcal V$, and looking at the action from two different coordinate systems, say $S$ and $S'$; see \textbf{Fig.\ref{fig:set1}}. Here is then the key idea:

{\it While it is a trivial statement that one can chose any coordinate system one likes, this fact can not be put to much use or gain any ``insight" if one uses an action which is not covariant. So we choose a generally covariant action.} 

The coordinate domains of $\mathcal V$ in $S$ and $S'$ ($ \DM^4 x$ and $ \DM^4 x'$) are, of course, different; call them $D$ and $D'$ respectively. We then have the following statement as trivially true:
\begin{widetext}
\begin{eqnarray}
0 = \int \limits_{\{\mathcal V, D'\}} L(\phi'(x'), \partial'_k \phi'(x'), g'_{ab}(x'), x'^i) \DM^4 x' - \int \limits_{\{\mathcal V, D\}}  L(\phi, \partial_k \phi, g_{ab}, x^i) \DM^4 x
\label{eq:zero-gen-cov}
\end{eqnarray}
\end{widetext}
where, by definition, 
\begin{eqnarray}
\phi'(x') &=& \phi(x)
\nonumber \\
g'_{ab}(x') &=& \frac{\partial x^i}{\partial x'^a} \frac{\partial x^j}{\partial x'^b} g_{ij}(x)
\end{eqnarray}
Although the zero above is trivial, one must realise exactly where the assumption of general covariance has gone in: the functional dependence
of $L$ on $\phi, g_{ab}$ remains the same in both coordinate systems. 

We now consider the coordinate transformation 
$$x'^k = x^k + \epsilon \xi^k(x)$$ 
where $\epsilon$ is a small parameter introduced solely for book-keeping (and has been set to unity wherever it is no longer needed), and use ordinary rules of integral calculus to change variables in the first integral above. We have 
\begin{eqnarray}
\frac{\partial x'^k}{\partial x^j} = \delta^{k}_{j} + \epsilon \partial_j \xi^{k}
\end{eqnarray}
which gives, for the Jacobian of the transformation,
\begin{eqnarray}
\mathrm{det} \l[ \frac{\partial x'^k}{\partial x^j} \r] = 1 + \epsilon ( \partial_k \xi^k ) + O(\epsilon^2)
\end{eqnarray}
Plugging this back into \eq{eq:zero-gen-cov}, we obtain
\begin{widetext}
\begin{eqnarray}
0 &=& \int_{\{\mathcal V, D\}} \Biggl( L(\phi'(x), \partial_k \phi'(x), g'_{ab}(x), x^i) + \epsilon \xi^k \partial_k L\Biggl) \Biggl( 1 + \epsilon ( \partial_k \xi^k )\Biggl) \DM^4 x - \int_{\{\mathcal V, D\}} L(\phi, \partial_k \phi, g_{ab}, x^i)  \DM^4 x
\nonumber \\
0 &=& \int_{\{\mathcal V, D\}} \Biggl( L(\phi', \partial_k \phi', g'_{ab}, x^i) - L(\phi, \partial_k \phi, g_{ab}, x^i) \Biggl) \DM^4 x +  \int_{\{\mathcal V, D\}}  \partial_k (L \xi^k) \DM^4 x
\nonumber \\
\end{eqnarray}
\end{widetext}
(Note the change from $D'$ to $D$ in the first integral.) The first two terms above can be obtained from the standard variation of the Lagrangian with respect to the variations 
\begin{eqnarray}
\delta \phi &=& \phi'(x)-\phi(x)=- \mathscr{L}_{\bm \xi} \phi 
\nonumber \\
\delta g_{ab} &=& g'_{ab}(x) - g_{ab}(x) = - \mathscr{L}_{\bm \xi} g_{ab}
\end{eqnarray}

\textit{The crucial point above is that we are now considering difference between quantities defined in the same coordinate system.} 

The result above can now be re-written in terms of $\mathcal{L} = L/\sqrt{-g}$:
\begin{widetext}
\begin{eqnarray}
0 &=& \int_{\{\mathcal V, D\}} \Biggl( \mathcal{L}(\phi', \partial_k \phi', g'_{ab}, x^i) - \mathcal{L}(\phi, \partial_k \phi, g_{ab}, x^i) \Biggl) \, \sqrt{-g}\, \DM^4 x +  
\int_{\{\mathcal V, D\}}  
\underbrace{\frac{1}{\sqrt{-g}} \partial_k ( \sqrt{-g} \mathcal{L} \xi^k)}_{\nabla_k (\mathcal{L} \xi^k)}
\, \sqrt{-g}\, \DM^4 x
\nonumber \\
\label{eq:passive-Final}
\end{eqnarray}
\end{widetext}
which is manifestly covariant, since the second integrand is just the expression for covariant divergence of the vector $\mathcal{L} \xi^k$.

We have therefore demonstrated how Lie derivatives appear naturally while dealing with a generally covariant action and considering the effects of coordinate transformation. Before we discuss further simplification of the above form, we will proceed to derive the same expression for active diffeomorphisms.

%
%
\subsection{The Active version: Diffeomorphisms}

We wish to repeat the above analysis by considering the change in action due to infinitesimal mapping of the points
in $\mathcal V$ to another physical region of the manifold, $\mathcal V'$; this is indicated in \textbf{Fig.\ref{fig:set1}}. To be specific, we shall consider the 
mapping to be $x^k \rightarrow x^k + \epsilon \xi^k (x)$. To start with, the action will undergo the obvious change under such
a transformation since one is now integrating the Lagrangian over a different physical region. This change is easily
computed as follows. Consider a small patch of the boundary of $\mathcal V$, with physical area $\DM A$ (if coordinates on the boundary are $y^\mu$, $\mu=1, 2, 3$, then $\DM A = \sqrt{h} \DM y^1 \DM y^2 \DM y^3$, where $h$ is the determinant of the metric induced on the boundary), and unit 
normal $\bm n$. Simple geometric considerations tell us that the volume swept by this patch of area is 
$\l( \bm \xi \bm \cdot \bm n  \r) \DM A$. Since the action is the volume integral of the Lagrangian, the corresponding
change in action is given by
\begin{eqnarray}
\delta S &=& \epsilon  \int_{\partial \mathcal V} \mathcal{L} \xi^k n_k \DM A 
\\
&=& \epsilon \int_{\mathcal V} \nabla_k \l( \mathcal{L} \xi^k \r) \sqrt{-g} \DM^4 x
\label{eq:action-var-Lxi}
\end{eqnarray}
where we have applied the divergence theorem.

We will now evaluate this same change in action by using the very same trick that we employed in the point particle case, by exploiting the general covariance of the action, and finally equate both the expressions. We keep ${\mathcal A}[\mathcal V]$ as it is, but choose to evaluate ${\mathcal A}[\mathcal V']$ in a different coordinate system. In particular, we wish to choose a coordinate system such that the coordinates of the points of $\mathcal V'$ in this new coordinate system become 
numerically equal to the coordinates of corresponding points in $\mathcal V$; i.e., symbolically, the point $x^i+\epsilon \xi^i$ in
$\mathcal V'$ is to be assigned new coordinates which are numerically equal to $x^i$. This will effectively map $\mathcal V' \to \mathcal V$, $D' \to D$.
As before, this can be done by the coordinate
transformation (note the difference from the passive version):
$$x'^k = x^k -  \epsilon \xi^k(x)$$ 
Written out explicitly
\begin{widetext}
\begin{eqnarray}
\delta \mathcal A\l[\phi\r] &=& 
\underbrace{\int \limits_{\{\mathcal V', D'\}} L(\phi(x), \partial_k \phi(x), g_{ab}(x), x^i) \DM^4 x}_{\mathrm{change~coordinates:~} x'^k = x^k - \xi^k(x)} 
\;\;\;-\;\;\;
\underbrace{\int \limits_{\{\mathcal V, D\}} L(\phi(x), \partial_k \phi(x), g_{ab}(x), x^i) \DM^4 x}_{\mathrm{keep~coordinates~unchanged}} 
\nonumber \\
\nonumber \\
\nonumber \\
&=& 
\int \limits_{\{\mathcal V, D\}} L(\phi'(x'), \partial'_k \phi'(x'), g'_{ab}(x'), x'^i) \DM^4 x'
\;\;\;-\;\;\;
\int \limits_{\{\mathcal V, D\}} L(\phi(x), \partial_k \phi(x), g_{ab}(x), x^i) \DM^4 x
\nonumber \\
&=& 
\underbrace{\int \limits_{\{\mathcal V, D\}} L(\phi'(x), \partial_k \phi'(x), g'_{ab}(x), x^i) \DM^4 x}_{\mathrm{labelling}~x'^k~\mathrm{as}~x^k} 
\;\;\;\;\;\;\;\;-\;\;\;
\int \limits_{\{\mathcal V, D\}} L(\phi(x), \partial_k \phi(x), g_{ab}(x), x^i) \DM^4 x
\nonumber \\
\label{eq:zero-active-diffeo1}
\end{eqnarray}
\end{widetext}
(Compare with the passive case, especially the arguments and the limits of integrations.) We now equate this to \eq{eq:action-var-Lxi} to obtain
\begin{widetext}
\begin{eqnarray}
 \int_{\mathcal V} \nabla_k \l( \mathcal{L} \xi^k \r) \sqrt{-g} \DM^4 x
&=& 
\underbrace{\int \limits_{\{\mathcal V, D\}} L(\phi'(x), \partial_k \phi'(x), g'_{ab}(x), x^i) \DM^4 x}_{\mathrm{labelling}~x'^k~\mathrm{as}~x^k} 
\nonumber \\
\;\;\;\;\;\;\;\;&-&\;\;\;
\int \limits_{\{\mathcal V, D\}} L(\phi(x), \partial_k \phi(x), g_{ab}(x), x^i) \DM^4 x
\nonumber
\label{eq:zero-active-diffeo2}
\end{eqnarray}
\end{widetext}
which we re-write as
\begin{widetext}
\begin{eqnarray}
0 &=& \int_{\{\mathcal V, D\}} \Biggl( \mathcal{L}(\phi', \partial_k \phi', g'_{ab}, x^i) - \mathcal{L}(\phi, \partial_k \phi, g_{ab}, x^i) \Biggl) \, \sqrt{-g}\, \DM^4 x -  
\int_{\{\mathcal V, D\}}  
\frac{1}{\sqrt{-g}} \partial_k ( \sqrt{-g} \mathcal{L} \xi^k)
\, \sqrt{-g}\, \DM^4 x
\nonumber \\
\label{eq:zero-active-diffeo3}
\end{eqnarray}
\end{widetext}
This is exactly the same as \eq{eq:passive-Final} once we recognise that here, since the coordinate transformation is $x'^k = x^k - \epsilon \xi^k(x)$,
\begin{eqnarray}
\delta \phi &=& \phi'(x)-\phi(x)=+\mathscr{L}_{\bm \xi} \phi 
\nonumber \\
\delta g_{ab} &=& g'_{ab}(x) - g_{ab}(x) = +\mathscr{L}_{\bm \xi} g_{ab}
\end{eqnarray}
thereby accounting for the overall minus sign. In fact, the various sign changes are perhaps the most important thing to note in the passive vs active versions discussed above. 
\section{Non-trivial results using Diffeomorphism Invariance}
So far, we have essentially highlighted the relation between general covariance, diffeomorphisms, and why Lie derivatives appear in the discussion of these. We now show that a proper understanding of diffeomorphisms can also provide a powerful tool to derive non-trivial results which otherwise would be derived using completely different methods, or are completely obscure as usually formulated. We will give one example of each:
	\begin{enumerate}
	\item Deriving equation of geodesic deviation using diffeomorphisms, keeping the curve fixed.
	\item Understanding the relation between canonical and metric stress-energy tensors in classical field theory.
	\end{enumerate}
\subsection{Point particles: Deriving the Geodesic deviation equation}

The equation of geodesic deviation is perhaps the most important equation in differential geometry since it directly describes the effects of curvature of space(-time) on behaviour of a congruence of test particles moving in that space(-time). More specifically, this equation gives the acceleration of the so-called deviation vector $\xi^i$ that connects any two members of such a congruence, and this acceleration turns out to depend directly on the Riemann curvature tensor of the background space(-time). Since curves play the main role in establishing this connection, we employ the ideas developed in this paper to re-derive the deviation acceleration in a novel manner, using diffeomorphisms.

\begin{figure}[!htb]
\begin{center}
\scalebox{0.35}{\includegraphics{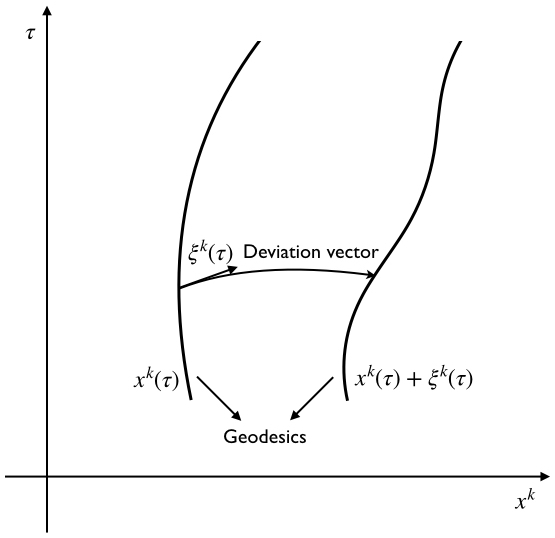}} 
\caption{The Geodesic deviation equation is one the most important equations in General Relativity. It gives the relative acceleration of two geodesics, which turns out to be a measure of space(time) curvature.}
\label{fig:geod-dev}
\end{center}
\end{figure}

The set-up here is two curves (with parameters synchronised), with tangent vectors $t^a$ and $t^a + \delta t^a$, and a deviation vector field $\xi^a(x(\tau))$ which measures the separation between them; see \textbf{Fig.\ref{fig:geod-dev}}. (At the lowest order, $\xi^a$ would simply be given by the coordinate differences of points on the two curves at the same parameter values, but this will not be true in general.)

For better clarity, we will keep things a bit more general and not assume the curves to be geodesics. The acceleration $a^i$ of these curves are then defined by \eq{eq:geodesic-eq}. Now suppose we wish to compute the difference $\delta a^i$ between the acceleration of the two curves (note that this is {\it not} the relative acceleration that we are interested in). We may use exactly the same trick here as we did while using active diffeomorphisms to ``vary without varying" the point particle action. 

We will now prove that this leads to
\begin{eqnarray}
\mathscr{L}_{\bm \xi} t^a &=& 0 
\nonumber \\
( \mathscr{L}_{\bm \xi} \Gamma^i_{\phantom{i}bc} ) \, t^b t^c &=& \mathscr{L}_{\bm \xi} a^i
\label{eq:lie-acc}
\end{eqnarray}

\textit{Proof}: Consider again \textbf{Fig.\ref{fig:geod-dev}}. Since acceleration $a^i$ is a vector, so will be it's difference, and one can compute this in any convenient coordinate system. In the spirit of ``varying without varying" introduced throughout previous sections, we make a coordinate transformation judiciously chosen such that, in the new coordinates, both the curves look the same. The only subtlety here is that, unlike the action, we are not dealing with a scalar obtained by integrating another scalar along a curve. Nevertheless, the algorithm remains the same. To simplify things, we may choose a coordinate transformation such that it is identity everywhere except in the small neighbourhood of the second curve. (The astute reader will quickly realise that we are just recreating the {\it hole} in Einstein's Hole argument.) In the new coordinates, then, the tangent vectors of both the curves will map to the same vector, which immediately yields the first equality above. However, the definition of $a^i$ also involves $\Gamma^a_{\phantom{a}bc}$, which will transform when coordinates are transformed. Hence, the only contribution to $\delta a^i$ comes from $\Gamma'^a_{\phantom{a}bc}-\Gamma^a_{\phantom{a}bc}$, immediately leading to \eq{eq:lie-acc}.

\textit{The end result, therefore, is the same as has been encountered repeatedly in earlier sections: The curve remains unchanged, every other geometric object ($g_{ab}, \Gamma^a_{\phantom{a}bc},$ etc.) changes by its Lie derivative.}

Note that the symbol $\mathscr{L}_{\bm \xi} t^a$ was also properly defined previously, and $\mathscr{L}_{\bm \xi} a^i$ must also be interpreted in a similar manner: it is the Lie derivative of acceleration vectors associated with a one parameter class of curves. 

We now use the results from Appendix \ref{app:Lie-derivs} to simplify the second expression above. 
\begin{eqnarray}
\l( \nabla_{(b} \nabla_{c)} \xi^i - R^{i}_{\phantom{a}(bc)k} \xi^k  \r) \, t^b t^c &=& \xi^k \nabla_k a^i - a^k \nabla_k \xi^i
\label{eq:geod-dev-active}
\end{eqnarray}
where we have used, for convenience, the covariant derivative, defined, for any vector $q^i$, as:
\begin{eqnarray}
\nabla_k q^i = \partial_k q^i +  \Gamma^i_{\phantom{i}km} q^m
\end{eqnarray}
Note that, with this definition, $a^i = t^k \nabla_k t^i$, and it can be verified that the covariant derivative satisfies all properties of ordinary derivatives, such as product rule. Moreover, whenever $\Gamma^i_{\phantom{i}km} = \Gamma^i_{\phantom{i}mk}$ (which is true in our case, see \eq{eq:geodesic-eq-active}), it is easily seen from the expressions given in Appendix \ref{app:Lie-derivs} that $\mathscr{L}_{\bm \xi} v^k =  \xi^i \nabla_i v^k -  v^i \nabla_i \xi^k$. Indeed, it can be proved in general that all the partial derivatives appearing in the computation of Lie derivatives can be replaced by covariant derivatives as long as $\Gamma^i_{\phantom{i}km}$'s are symmetric in lower indices.

Now consider,
\begin{eqnarray}
t^b t^c \, \nabla_{(b} \nabla_{c)} \xi^i &=& t^b t^c \, \nabla_{b} \nabla_{c} \xi^i
\nonumber \\
&=& t^b \nabla_{b} \l( t^c \nabla_{c} \xi^i \r) - a^c \nabla_{c} \xi^i
\end{eqnarray}
Plugging this is \eq{eq:geod-dev-active}, we obtain
\begin{eqnarray}
 t^b \nabla_{b} \l( t^c \nabla_{c} \xi^i \r) = R^{i}_{\phantom{a}(bc)k} \xi^k \, t^b t^c + \xi^k \nabla_k a^i
\label{eq:geod-dev-active2}
\end{eqnarray}
For $a^i=0$ (geodesics), this is precisely the geodesic deviation equation.

\subsection{Field theory: Relation between Canonical and metric stress-energy tensors}
Our next example is from field theory, and we here wish to use the results of {\bf Sec III} to prove a relation between the so called canonical and metric stress-energy tensors of a scalar field. These stress-energy tensors are derived using two very different definitions, and standard texts do not discuss the relation between these. We will now show that such a relation is, in fact, a direct consequence of diffeomorphism invariance of the action. 

Both the active and passive versions of diffeomorphisms we discussed above, applied to arbitrary spacetime regions, essentially imply:
\begin{eqnarray}
{L}(\phi', \partial_k \phi', g'^{ab}, x^i) - {L}(\phi, \partial_k \phi, g^{ab}, x^i) - \partial_k ( L \xi^k) = 0
\nonumber
\end{eqnarray}
where, for convenience, we used $g^{ab}$ - the inverse of $g_{ab}$ - as the basic variable, and recalled our definition $L = \mathcal{L} \sqrt{-g}$. In the above, 
$\phi'(x)=\phi(x) + \mathscr{L}_{\bm \xi} \phi$ and $g'^{ab}(x)=g^{ab}(x) + \mathscr{L}_{\bm \xi} g^{ab}$. We will now work on this identity to obtain conditions imposed by diffeomorphism invariance, analogous to the ones we obtained for reparameterisation invariance of point particle action, viz. \eq{eq:Eeq0}. 

Applying chain rule, we first obtain
$$
\frac{\partial L}{\partial x^i} = \frac{\partial L}{\partial \phi} \partial_i \phi + \frac{\partial L}{\partial (\partial_k \phi)} \partial_i \partial_k \phi + 
\frac{\partial L}{\partial g^{ab}} \partial_i g^{ab}  +
\frac{\delbar L}{\partial x^i}
$$
where ${\delbar L}/{\partial x^i}$ simply means partial derivative of ${L}(\phi, \partial_k \phi, g^{ab}, x^i)$ with respect to the last argument $x^i$, keeping everything else fixed. In contrast, ${\partial L}/{\partial x^i}$ is simply the gradient of $L$ treated as the function of coordinates. Further, since $g^{ab}$ is symmetric, we will take ${\partial L}/{\partial g^{ab}}$ to be symmetric as well. With all this in hand, we set to work on our identity

\begin{eqnarray}
0  &=& {L}(\phi', \partial_k \phi', g'^{ab}, x^i) - {L}(\phi, \partial_k \phi, g^{ab}, x^i) - \partial_k ( L \xi^k)
\nonumber \\
\nonumber \\
0  &=& 
\frac{\partial L}{\partial \phi} \mathscr{L}_{\bm \xi} \phi + \frac{\partial L}{\partial (\partial_k \phi)} \partial_k (\mathscr{L}_{\bm \xi} \phi) + 
\frac{\partial L}{ \partial g^{ab}} \mathscr{L}_{\bm \xi} g^{ab} - \partial_k ( L \xi^k)
\nonumber \\
\nonumber \\
0  &=&  
\xi^i \l( \frac{\partial L}{\partial x^i} - \frac{\delbar L}{\partial x^i} \r) + \frac{\partial L}{\partial (\partial_k \phi)} (\partial_i \phi) \, (\partial_k \xi^i)
+ \frac{\partial L}{\partial g^{ab}} \l( \mathscr{L}_{\bm \xi} g^{ab} - \xi^i \partial_i g^{ab} \r)
 -  \frac{\partial}{\partial x^i} \l(L \xi^i \r)
\nonumber \\
\nonumber \\
0  &=&  
- \xi^i \frac{\delbar L}{\partial x^i} 
- \l( - \frac{\partial L}{\partial (\partial_k \phi)} (\partial_i \phi) + L \delta^k_i\r) \, (\partial_k \xi^i)
+ \frac{\partial L}{\partial g^{ab}} \l( \mathscr{L}_{\bm \xi} g^{ab} - \xi^i \partial_i g^{ab} \r)
\nonumber \\
\nonumber \\
0  &=&  
- \xi^i \frac{\delbar L}{\partial x^i} 
- \l( - \frac{\partial L}{\partial (\partial_k \phi)} (\partial_i \phi) + L \delta^k_i\r) \, (\partial_k \xi^i)
-2 \frac{\partial L}{\partial g^{ab}} \l( g^{ka} \partial_k \xi^b \r)
\nonumber \\
\nonumber \\
0  &=&  
- \xi^i \frac{\delbar \mathcal{L}}{\partial x^i} 
- \Biggl[ \underbrace{- \frac{\partial \mathcal{L}}{\partial (\partial_k \phi)} (\partial_i \phi) + \mathcal{L} \delta^k_i}_{t^k_{\phantom{k}i}} \Biggl] \, (\partial_k \xi^i)
+\Biggl[ \underbrace{- \frac{2}{\sqrt{-g}} \frac{\partial ( \mathcal{L} \sqrt{-g})}{\partial g^{ai}} g^{ka}}_{T^k_{\phantom{k}i}} \Biggl] \partial_k \xi^i
\nonumber \\
\nonumber \\
0  &=&  
- \xi^i \frac{\delbar \mathcal{L}}{\partial x^i} 
- \Biggl[ t^k_{\phantom{k}i} -T^k_{\phantom{k}i} \Biggl] \partial_k \xi^i
\end{eqnarray}
where the third step follows after a few lines of simple and straightforward algebra, using the chain rule stated above, and in the last step we have re-introduced 
$\mathcal{L}$. Note that, by definition, $\delbar$ will not act on $\sqrt{-g}$, since it only picks up the {\it explicit} dependence of Lagrangian on $x^i$, {\it keeping the metric and the field fixed}. 

The result above is the precise analog of Eq.~(\ref{Eeq0-deriv}) for point particle action. Since one may choose $\xi^i$ and its derivatives arbitrarily at any point in the domain, we may read-off the analogue of \eq{eq:Eeq0}:
\boxedeq{
t^k_{\phantom{k}i} -T^k_{\phantom{k}i}=0 \hspace{.5cm} {\rm and}  \hspace{.5cm}  \frac{\delbar \mathcal{L}}{\partial x^i}=0
}{Tab-eq-0}

The second identity is, again, straightforward to understand (and could have been guessed). The first one is interesting. The object $t^k_{\phantom{k}i}$ is what one calls in standard field theory as the canonical stress-energy tensor. The object $T^k_{\phantom{k}i}$, on the other hand, is the stress-energy tensor one encounters in general relativity by varying the action with respect to the metric tensor. The first condition then essentially equates the two stress tensors. This is known to be true for conventional scalar field actions. In fact, we leave the reader with the following

\textbf{\textit{Homework}:} Repeat the above analysis for a vector field theory described by the Lagrangian (density) $\mathcal{L}(A_{i},\partial_{m}A_{k},g^{ab},x^{i})$, and show that:
\begin{eqnarray}
	\begin{aligned}
		0 = - \xi^i \frac{\delbar \mathcal{L}}{\partial x^i} 
		&- \left(t^{k}_{\phantom{k}i} - \pi^{j(k)} \partial_{j}A_{i} - T^{k}_{\phantom{k}i}\right)(\partial_{k}\xi^{i})
		\nonumber \\
		& \hspace{1.25cm} + \left( \frac{\partial\mathcal{L}}{\partial A_{k}}\partial_{k}\xi^{i}+\pi^{mk}\partial_{m}\partial_{k}\xi^{i} \right)A_{i}
	\end{aligned}
\end{eqnarray}
where $\pi^{m(k)}={\partial\mathcal{L}}/{\partial(\partial_{m}A_{k})}$ is the momentum conjugate to $A_i$. Interpret the conditions arising from this for the case of electromagnetic Lagrangian: $\mathcal{L}=-(1/4) F_{ab} F^{ab}$, and verify that, once again, you obtain the correct identity between canonical and metric stress tensor. For the more general case, since $\partial_i A_k$ is not covariant, to get the correct definition for conjugate momentum, one must introduce the connection 
$\Gamma^a_{\phantom{a}bc}$ into the Lagrangian, and consider, instead, $\mathcal{L}(A_{i},\partial_{m}A_{k},g^{ab}, \Gamma^a_{\phantom{a}bc}, x^{i})$. Try this out and interpret the resultant conditions. For getting started, you may use the computational tools described in \cite{ajp-field-theory}.

\section{Discussion}
It is well known that writing an action functional in a generally covariant manner allows one to deduce important identities. For example, in general relativity, where the relevant field is the metric tensor $g_{ab}(x)$ itself, described by the Einstein-Hilbert lagrangian ${\mathcal L} = (16 \pi G)^{-1} R$ (in units where the speed of light is one), where $R$ is the Ricci scalar, our argument, with suitable choice of the variation vector field $\xi^i(x)$, gives the well-known (contracted) Bianchi identities, $\nabla_a G^a_b = 0$, for the Einstein tensor $G^a_{b}$. Indeed, several textbooks on GR do discuss this argument. However, one may derive even stronger identities without putting any constraint on $\xi^i$. For example, 
one may show that 
$$
\nabla_a \l[ 2 G^a_b \xi^b + \delta_\xi v^a + R \xi^a \r] = 0
$$
where $\delta_\xi v^a$ is the surface term that appears in the variation of the action. (More precisely, the $\delta_\xi v^a$ term arises from the $g^{ab} \delta R_{ab}$ part in the variation of Einstein-Hilbert action, and can be found in any standard text on general relativity. It's precise form is not relevant for the point we are trying to make, though it is easy to see that $\nabla_a  \delta_\xi v^a = - 2 \nabla_a (R^a_b \xi^b)$). This identity - which has the form of a conservation law - could not have been guessed without considerable effort, particularly when we move to more general theories of gravity where similar identities will hold.

Of course, one could have chosen to {\it not} formulate the action in a generally
covariant manner (in general relativity, the so-called $\Gamma$-$\Gamma$ Lagrangian provides an example) and yet obtain the same field equations, 
since what matters in deriving the field equations is the \textit{variation} of the action and not the action itself. Of course, the same identities would still exist, except that figuring them out would be much more non-trivial. I hope the above paper highlights the importance of formulating an action in a generally covariant, by showing that doing so allows one to establish certain identities quite easily. I also hope that the analysis presented here would bring out clearly the true significance of Lie derivatives of tensor fields, and why such variations are related to gauge degrees of freedom in gravitational theories. 

\section*{Acknowledgements}
I thank Hari K. for reading through the manuscript and providing useful comments. I am also very grateful to the two referees for their very detailed, thoughtful, and insightful comments as well as criticisms that have helped in improving the presentation in certain advanced sections of the paper, and thereby broaden its pedagogical scope. 

\appendix
\section{Lie derivatives} \label{app:Lie-derivs}
In this Appendix, we recall the relevant expressions for what are called Lie derivatives. To avoid unnecessary digression, we will be brief refer the reader to \cite{lie-derivs} for more detailed discussion of Lie derivatives. In any case, one of the key aims of this paper is to indicate how Lie derivatives arise naturally in variations associated with a generally covariant action, so the reader should be able to connect their definition with the geometric set-up of the main text.

Consider a vector field $v^k(x)$. Under a coordinate transformation to $x'^i=x^i+\epsilon \xi^i(x)$ (where $\epsilon$ is a book-keeping parameter and $\xi^i(x)$ a given vector field), this transforms as 
\begin{eqnarray}
{v'}^k(x') &=& \frac{\partial x'^k}{\partial x^i} v^i(x)
\nonumber \\
&=& v^k(x) + \epsilon (\partial_i \xi^k) v^i(x) + O(\epsilon^2)
\nonumber
\end{eqnarray}
The Lie derivative is then defined by the difference 
\begin{eqnarray}
\mathscr{L}_{\bm \xi} v^k &\overset{\rm def}{:=}& - \lim \limits_{\epsilon \to 0} \frac{1}{\epsilon} \l(  v'^k(x^i) - v^k(x^i) \r)
\nonumber \\
&=& - \lim \limits_{\epsilon \to 0} \frac{1}{\epsilon} \l(  v'^k(x'^i-\epsilon \xi^i) - v^k(x^i) \r)
\nonumber \\
&=& - \lim \limits_{\epsilon \to 0} \frac{1}{\epsilon} \l(  v'^k(x'^i) - \epsilon \xi^i \partial_i v^k - v^k(x) + O(\epsilon^2) \r)
\nonumber \\
&=& - \lim \limits_{\epsilon \to 0} \frac{1}{\epsilon} \l( \epsilon (\partial_i \xi^k) v^i(x) - \epsilon \xi^i \partial_i v^k + O(\epsilon^2) \r)
\nonumber \\
&=& \xi^i \partial_i v^k -  v^i \partial_i \xi^k
\label{eq:lie-deriv-vec-field}
\end{eqnarray}
where, in the 3rd equality, we have used the fact that ${v'}^k(x') = v^k(x) + O(\epsilon)$. One may follow exactly these same set of steps 
for tensors of arbitrary ranks in a straightforward manner. In fact, one can even compute Lie derivatives of objects which are not tensors (such as 
$ \Gamma^a_{\phantom{a}bc}$), since all that is required is the knowledge of how these objects transform under coordinate transformations. Below, we give a list of Lie derivatives of some important geometrical objects that are used in the main text.
\begin{eqnarray}
\mathscr{L}_{\bm \xi} \phi(x) &=& \xi^i \partial_i \phi
\nonumber \\
\mathscr{L}_{\bm \xi} g_{ab}(x) &=& \xi^i \partial_i g_{ab} + g_{am} \partial_b \xi^m + g_{mb} \partial_a \xi^m
\nonumber \\
\mathscr{L}_{\bm \xi} g^{ab}(x) &=& \xi^i \partial_i g^{ab} - g^{am} \partial_m \xi^b - g^{mb} \partial_m \xi^a
\nonumber \\
\mathscr{L}_{\bm \xi} \Gamma^a_{\phantom{a}bc} &=& \nabla_{(b} \nabla_{c)} \xi^a - R^{a}_{\phantom{a}(bc)k} \xi^k 
\end{eqnarray}
where $Q_{(ab)} := (Q_{ab}+Q_{ba})/2$, and the Christoffel symbols $\Gamma^a_{\phantom{a}bc}$ were introduced in \eq{eq:geodesic-eq-active}, from which its Lie derivative can be computed by using $\mathscr{L}_{\bm \xi} g_{ab}(x)$. This is a straightforward, though cumbersome, computation, but the result is very interesting due to the appearance of the tensor $R^{a}_{\phantom{a}(bc)k}$, which reads:
$$
R^{a}_{\phantom{a}bcd} = \partial_c \Gamma^a_{\phantom{a}bd} - \partial_d \Gamma^a_{\phantom{a}bc} 
+ \Gamma^a_{\phantom{a}kc} \Gamma^k_{\phantom{k}bd}  - \Gamma^a_{\phantom{a}kd} \Gamma^k_{\phantom{k}bc}   
$$
and is known as the Riemann curvature tensor - the single most important tensor that carries all the information about the curvature of space(time).
\\
\\
The above purely mathematical definition of Lie derivatives obscures their true geometric relevance. One of the main themes behind many results discussed in this paper is, in fact, to clearly explain how Lie derivatives help formalise the idea of \textit{using together the flow under a vector field and freedom of coordinate transformations} to define notion of a derivative, and how that connects them to diffeomorphism invariance and general covariance. If you think a bit deeper about all the examples along with the above definitions, Lie derivatives provide a notion of derivative in which the change in a tensor field as one goes from a point $p$ to $q$ along a given vector field $\xi^k$ is evaluated after also simultaneously doing a coordinate transformation such that \textit{numerical values of coordinates} of $q$ in the new coordinate system are same as those of $p$ in the old coordinate system. Obviously, such a derivative has to be deeply connected with transformation properties of tensors under coordinate transformations.

\widetext


\end{document}